\documentclass[runningheads]{llncs}
\usepackage{amssymb,amsmath,array}
\usepackage{makecell}
\usepackage{booktabs}
\usepackage{hyperref} 

\usepackage[T1]{fontenc}
 
\usepackage[misc]{ifsym} 

\usepackage{graphicx}

\usepackage{orcidlink}

\begin{document}

\title{Real-time and personalized product recommendations for large e-commerce platforms}
\titlerunning{Real-time and personalized product recommendations}

\author{
Matteo Tolloso\orcidlink{0009-0002-1025-7363}\inst{1}\Letter\thanks{Corresponding author: \email{matteo.tolloso@phd.unipi.it}} \and
Davide Bacciu\orcidlink{0000-0001-5213-2468}\inst{1} \and
Shahab Mokarizadeh\orcidlink{0000-0002-9292-3552}\inst{2} \and
Marco Varesi\inst{2}
}

\authorrunning{M. Tolloso et al.}

\institute{
University of Pisa - Department of Computer Science \and
H\&M - AI, Analytics \& Data
}

\maketitle    

\begin{center}
\textit{\small © 2025. This is the author's version of a work that was accepted for publication in the proceedings of the International Conference on Artificial Neural Networks (ICANN) 2025. The final authenticated version will be published by Springer in the Lecture Notes in Computer Science (LNCS) series.}
\end{center}

\begin{abstract}
We present a methodology to provide real-time and personalized product recommendations for large e-commerce platforms, specifically focusing on fashion retail. Our approach aims to achieve accurate and scalable recommendations with minimal response times, ensuring user satisfaction, leveraging Graph Neural Networks and parsimonious learning methodologies. Extensive experimentation with datasets from one of the largest e-commerce platforms demonstrates the effectiveness of our approach in forecasting purchase sequences and handling multi-interaction scenarios, achieving efficient personalized recommendations under real-world constraints.

\keywords{Recommendation Systems  \and Graph Neural Networks \and Parsimonious Learning.}
\end{abstract}

\section{Introduction}
Recommendation systems support users in their discovery of products, especially in online retail, through integration of preferences, behaviors, and trends. In digital marketplaces aspects such as personalization and responsiveness are paramount to maintain a competitive edge \cite{de2024personalized}. These requirements pose strong methodological and technological challenges on the learning and predictive substrate underlying the recommendation system. On the one hand, the learning model should be able to dynamically and swiftly update to respond both to changing behaviors and trends at a population level, as well as to personalize recommendations to the (possibly short-term) history of the specific user. At the same time, aspects of responsiveness and efficiency cannot be overlooked to avoid affecting the user experience on the digital platform.  

With such motivations in mind, we propose a novel graph-based recommendation system \cite{gao2023survey} designed purposely for a fashion item recommendation scenario, characterized by the large scale of the item repository (and hence the graph) and tight needs of few-shot-like personalization of the recommendation to the personal taste and stock availability.
Our framework is designed to target, specifically: (i) near real-time recommendations during e-commerce navigation with response times in the order of few milliseconds; (ii) scalability with respect to the number of users and reduction of cold-start effects; (iii) continual recommendation adaptation based on stock availability, with a maximum delay of a few minutes; (iv) personalized recommendations integrating historical and current session engagement; (v) efficient continual training limiting resource usage to few GPU hours per week. 

Our approach builds upon the use of Graph Neural Networks (GNNs) \cite{bacciu2020gentle} to process articulated user-item interaction patterns within real-world recommendation systems \cite{gao2023survey,dukic2022inductive}. We specifically tackle the open-problem of working with multi-relational graphs  \cite{sattar2024multi} in a continuous update scenario in large-scale deployments. Such a challenging scenario requires integrating powerful GNN models, that can effectively extract information from complex multi-relational graphs, with knowledge distillation \cite{gou2021knowledge}, which enables the transfer of knowledge from the large GNN teacher to smaller and efficient personalized student models. This is complemented with the use of few-shot \cite{song2023comprehensive} and continual learning \cite{wang2024comprehensive} methodologies to allow efficient adaptation to new data, with minimal use of samples and without forgetting past knowledge. 

The key contributions of this work are as follows:
\begin{itemize}
    \item We introduce a novel distillation framework where the GNN is used to produce on-the-fly efficient personalized models capable of providing user-tailored recommendations in the order of 1ms (on CPU) and with limited memory fingerprint (700kb).
    \item We introduce an efficient approach to continually personalize user recommendation models, requiring on average less than 100ms to complete adaptation (on CPU).
    \item We provide an empirical validation of the effectiveness
of our approach in two large-scale benchmarks using real-world data (one publicly available, one proprietary).
\end{itemize}

\section{Parsimonious continual adaptation of personalized recommendation models}
\label{sec:methods}

We build on a graph-based representation of historical product data, where items are assumed to be associated with visual information describing its nature (i.e. a picture of the product).  In particular, we consider a heterogeneous graph $\mathcal{G}$ where nodes correspond to products, while edges mark the fact that users interacted with both products. Given an edge between two products (nodes), $p$ and $q$, this edge will be associated with a weight, $a_{p,q}^{c_{i}}$, that quantifies the intensity of the interaction between these products, i.e. the number of users that interacted with both products. Edges are also associated with a type $c_i$ that corresponds to one of the following four interaction events: \textit{co-clicked}, \textit{co-favorite}, \textit{co-cart}, and \textit{co-purchased}. Each node $p$ has attached an initial label $h_p^0$ that is a vectorial embedding of the product image, obtained through the pre-trained Convolutional Neural Network (CNN) Resnet-18 \cite{he2016deep}.

\begin{figure}[h!]
\centering
\includegraphics[width=\linewidth]{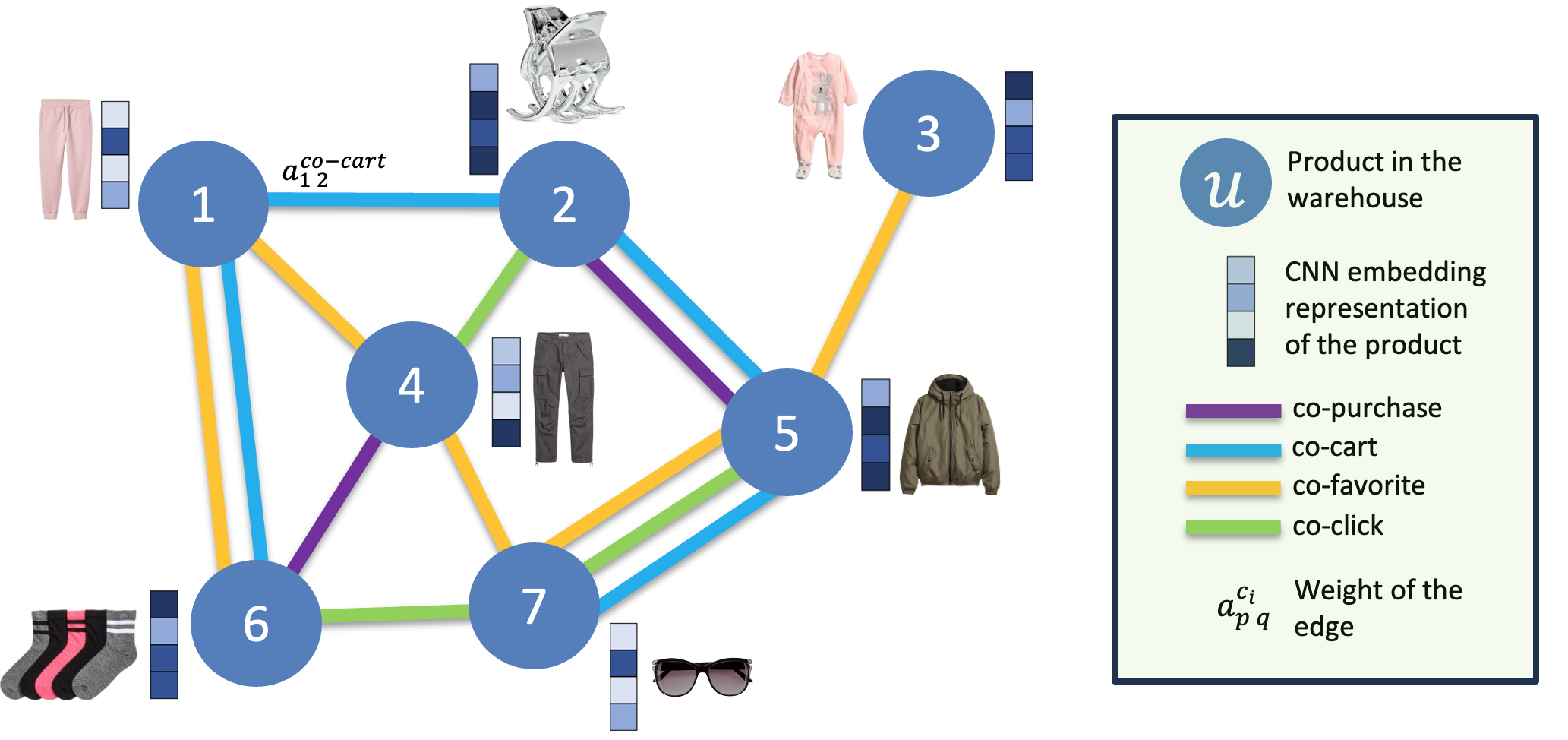}
\caption{Simplified example of a historical interaction graph used in our recommendation system. Nodes represent individual fashion products, each associated with visual embeddings derived from product images. Edges between nodes signify user interactions and are categorized into four interaction types: \textit{co-clicked}, \textit{co-favorite}, \textit{co-cart}, and \textit{co-purchased}. Each edge carries a weight proportional to the number of users who performed that specific interaction between the two products. The resulting heterogeneous graph encodes complex relationships among products, enabling the system to learn interaction patterns for personalized recommendations.}
\label{fig:graph}
\end{figure}

Figure \ref{fig:graph} shows an illustrative example of the historical heterogeneous graph.

\begin{figure}[h!]
\centering
\includegraphics[width=\linewidth]{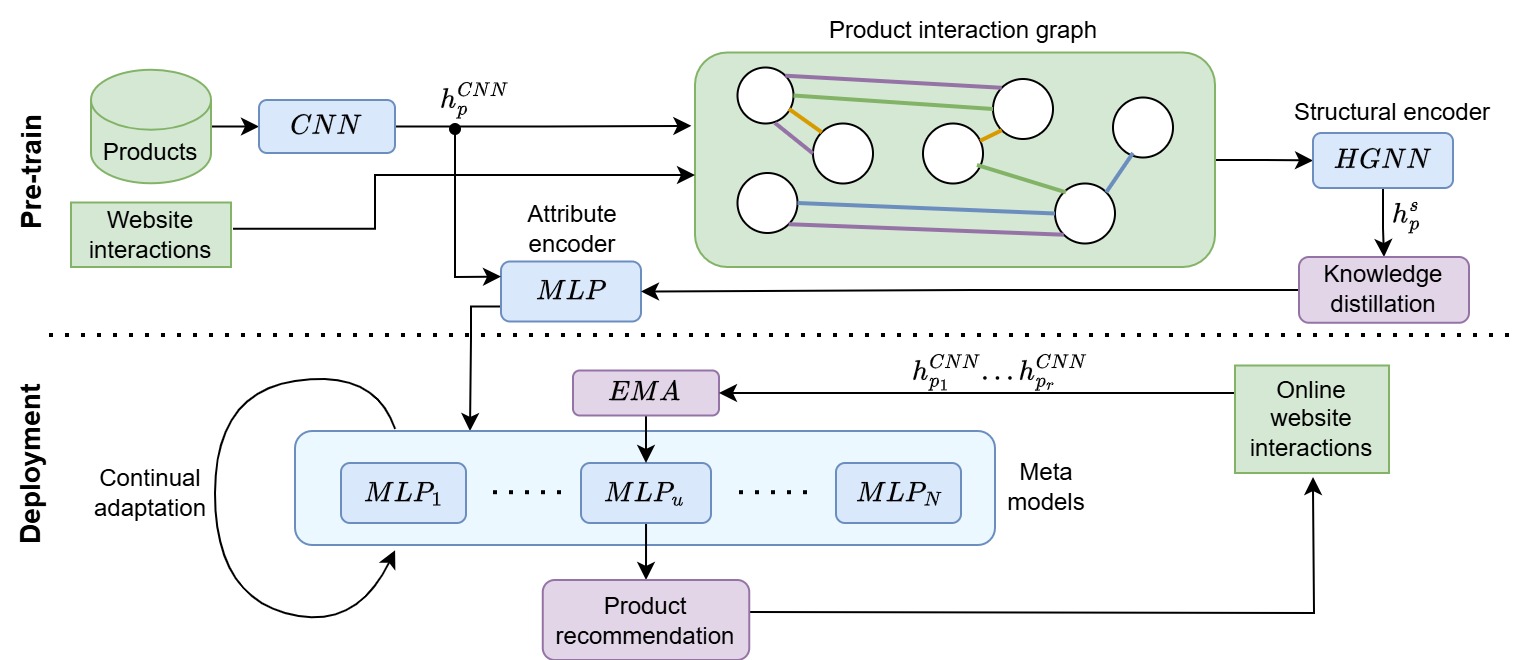}
\caption{Overall architecture of our proposed recommendation framework, consisting of three core components: (i) the \textit{structural encoder (HGNN)}, which uses a Heterogeneous Graph Neural Network to process the historical interaction graph and capture intricate relational information between products; (ii) the \textit{attribute encoder (Student MLP)}, a lightweight Multi-Layer Perceptron trained via knowledge distillation to approximate the HGNN embedding space using only product image embeddings; and (iii) the \textit{personalized meta-model (Personal MLP)}, an individual user-specific version of the attribute encoder, continually fine-tuned on recent user interactions to personalize the embedding space. The system generates real-time product recommendations by efficiently finding the nearest neighbors to the dynamically updated user embedding.}
\label{fig:overall}
\end{figure}

The overall architecture (Figure \ref{fig:overall}) of our recommendation system comprises three components: (i) a \textit{structural encoder}, (ii) an \textit{attribute encoder}, and (iii) a personalized \textit{meta model} for each user. The \textit{structural encoder} processes the historical product graph using a Heterogeneous Graph Neural Network (HGNN)\cite{zhang2019heterogeneous}. 
A Graph Neural Network (GNN) \cite{bacciu2020gentle} is a machine learning architecture designed to handle graph-structured data. It employs a feedforward approach where each layer $l$ generates node representation for a node $p$, named $h_p^l$, by aggregating information from neighboring nodes according to:
\begin{equation}
	{h}_p^{l} = \Phi^{l} \left( {h}_p^{l-1}, \Psi \left( \left\{ \Omega^{l}({h}_q^{l-1}) \mid q \in {N}_p \right\} \right) \right).
	\label{eq:gnn}
\end{equation}
In this formula, $N_p$ is the set of neighbors of node $p$, their representation in the layer $l-1$ is projected into a new space via the function $\Omega$ and aggregated into a single neighborhood representation using a permutation-invariant function $\Psi$ (such as max, mean, or sum). Finally, the function $\Phi$ combines this aggregated neighborhood representation with the embedding of node $p$ to produce the updated representation. Typically, $\Omega$ and $\Phi$ are Multi-Layer Perceptrons and the initial representation of a node $p$ $h_p^0$ is the initial feature vector associated with the node.
When dealing with heterogeneous graphs, a simple but effective approach to build an HGNN is to use multiple GNNs, one for each homogeneous subgraph within the heterogeneous graph, and combine the multiple representations of a node $p$ in each layer $l$ \cite{zhang2019heterogeneous}. The aggregation function is usually the mean, but more complex approaches exist \cite{sattar2024multi}.

The HGNN is trained via self-supervision similarly to \cite{hao2020inductive}. The model learns to project the original product $p$ representation $h_p^{CNN}$ (obtained from the CNN) in an embedding space where similarly interacted articles are closer. For each node $p$ the HGNN produces a structural embedding $h_p^s = HGNN(h_p^{CNN}, E_+)$, where $E_+$ are the edges of the graph, and its training is guided by the contrastive loss 

\begin{align}
	\label{eq:structural_encoder_loss}
	\mathcal{L}_{cont}^{c_i} &= \frac{1}{|E_+^{c_i}|} \sum_{(p, q)\in E_+^{c_i}} a_{p,q}^{c_i}||h_p^s - h_q^s ||_2^2 
	-  \frac{1}{|E_-^{c_i}|}\sum_{(p, q) \in E_-^{c_i}} ||h_p^s - h_q^s ||_2^2,
\end{align}
that is replicated for each different relation type $c_i$ in the graph. Here, $E_-^{c_i}$ are negative edge samples (edges that are not in the graph) selected randomly with $|E_-^{c_i}| = |E_+^{c_i}|$, and $a_{p, q}^{c_i}$ is the weight of the edge between the nodes $p$ and $q$. The final loss is:
\begin{equation}
L_{cont}^{tot} = \gamma_1 \times L_{cont}^{c_1} + \gamma_2 \times L_{cont}^{c_2} + \gamma_3 \times L_{cont}^{c_3} + \gamma_4 \times L_{cont}^{c_4},
\end{equation}
with $\gamma_i$ hyper-parameters to assign importance to the different relation types $c_i$.

\begin{figure}[h!]
    \centering
    \includegraphics[width=\linewidth]{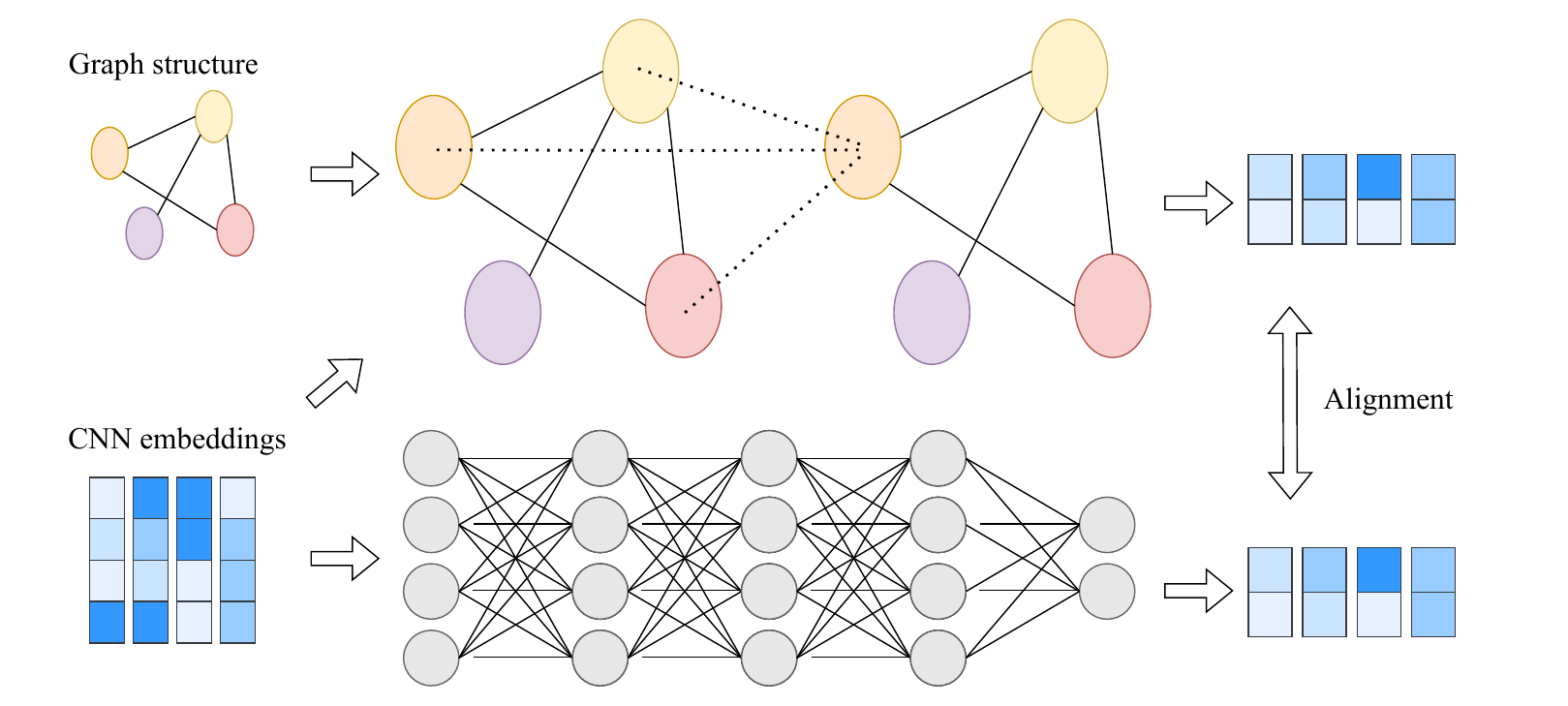}
    \caption{Distillation process. On the upper part the GNN takes in input the CNN embeddings and the graph structure. On the bottom part the MLP takes in input only the CNN embeddings. The alignment loss pushes the two embeddings spaces to be similar. Note that the embeddings' dimensionality is also compressed by the two networks. }
    \label{fig:distillation}
\end{figure}

The HGNN serves as a teacher model to transfer knowledge to the lightweight student model, i.e. the \textit{attribute encoder} (Figure \ref{fig:distillation}). The attribute encoder is a simple Multi-Layer Perceptron (MLP) that learns to map the original CNN embedding of the product into the embedding space generated by the HGNN (and that takes into consideration the relational inter-product knowledge) without having access to neighborhood information. From a recommendation systems perspective, such a student model is performing \textit{hybrid filtering} \cite{gao2023survey} by integrating \textit{content-based filtering} (the CNN embeddings) with \textit{collaborative filtering} (the HGNN embeddings). Training of the student model minimizes the alignment loss
\begin{equation}
	\label{loss_align}
	\mathcal{L}_{al} = \frac{1}{|V|} \sum_{p \in V} ||MLP(h_p^{CNN}) - HGNN(h_p^{CNN}, E_+)||_2^2,
\end{equation}
where $MLP(h_p^{CNN})$ is the embedding of the node $p$ produced by the attribute encoder and $V$ is the set of all the nodes in the graph.

The attribute encoder is then the model upon which we build the \textit{meta-model} \cite{vettoruzzo2024advances} for continuous personalization. Specifically, when a new user enters the shopping platform it is instantiated a personal copy of the attribute encoder (the meta-model). These personal MLPs project the products into a space that is continually adapted based on personal interactions with the website, and whose product embeddings are ultimately used to provide the personalized recommendations. 

The recommendation process is very efficient: a user's representation is kept in memory as an Exponential Moving Average (EMA) of the user's interacted articles as 
\begin{equation}
    u_t = (1 - \alpha) \times u_{t-1} + \alpha \times MLP_{u}(h_{t, u}^{CNN}), u_0 = MLP_{u}(h_{0, u}^{CNN}),
\end{equation}
where $h_{t, u}^{CNN}$ is the CNN embedding of the product interacted by the user $u$ at time $t$ and $MLP_{u}(h_{t, u}^{CNN})$ is its projection made by the personal MLP of the user $u$.
The $K$-nearest-neighbors to $u_t$, in the space projected by $MLP_u$, are the selected suggestions.

\begin{figure}[h!]
\centering
\includegraphics[width=0.7\linewidth]{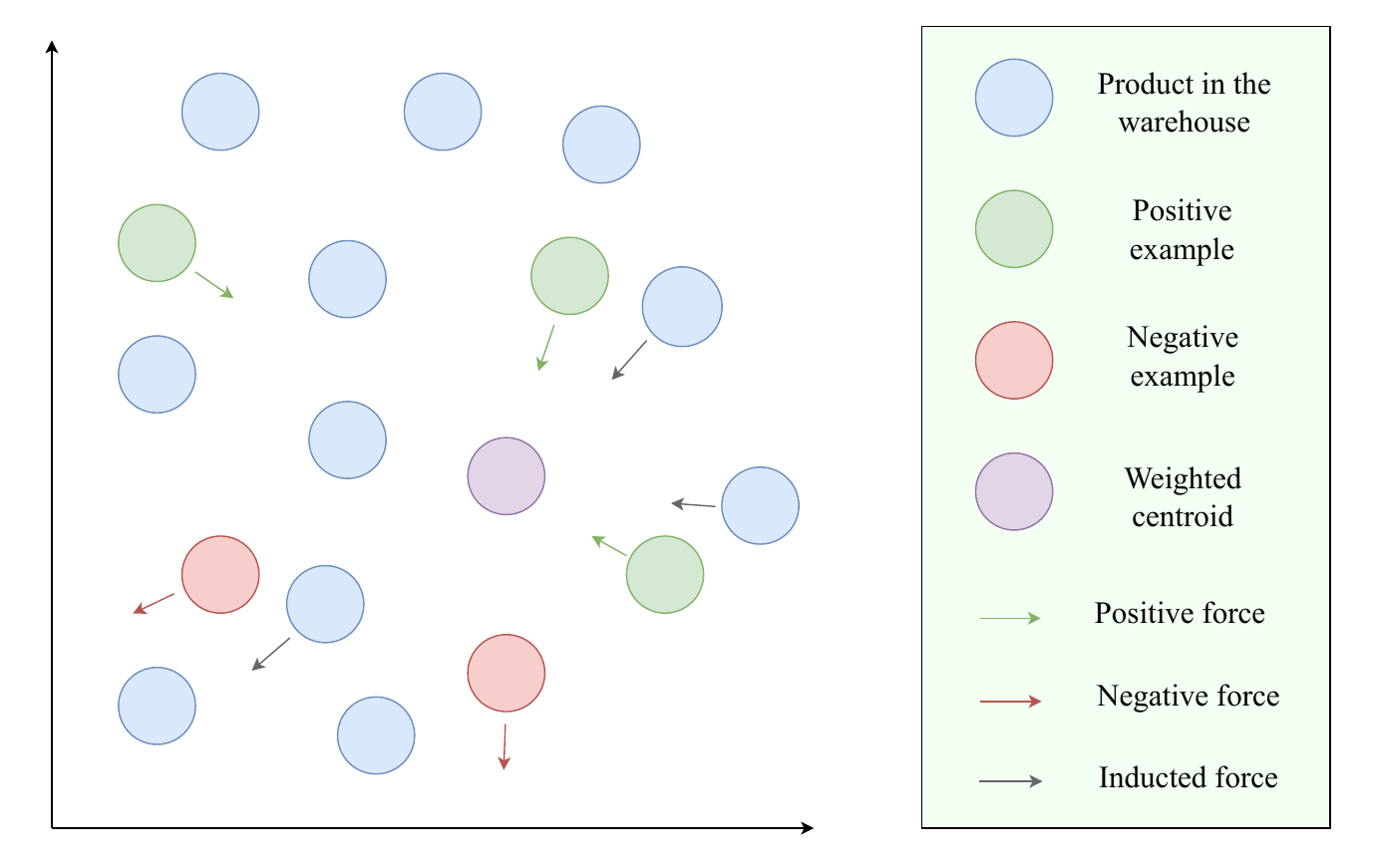}
\caption{Continual personalization process within the recommendation framework. When a user interacts with products (green points), the user's personal embedding space is adjusted so that these positively interacted items move closer to a calculated user preference centroid. Consequently, related yet unexplored products (nearby gray points) also shift within this personalized space, enhancing the likelihood that they become future recommendations. Products not interacted with (red points) serve as negative examples, guiding the model to maintain sufficient separation in the embedding space. This dynamic adjustment leverages recent user behavior to continuously refine personalized recommendations.}
\label{fig:personalization}
\end{figure}

To adapt the personal embeddings of user $u$, we collect a batch of interactions $B_u$ and we calculate the weighted centroid $h^{wc}_u$ of the CNN embeddings of interacted products $h_{p+}^{CNN}$ as:
\begin{equation}
   h^{wc}_u = \frac{1}{|B_u|}\sum_{(h_{p+}^{CNN}, w) \in B_u} h_{p+}^{CNN} \times w , 
\end{equation} 
where $w$ weighs the importance of the interaction (for example purchase: 4, cart: 3, favorite: 2, click: 1). The centroid represents the current interest of the user. The next step brings positively interacted articles closer to the centroid and consequentially closer together. The hypothesis is that we are modifying the whole space in a way such that the related products, for the user's personal preference, are moved closer together. We also select negative examples $h_{p-}^{CNN}$ among the products that the system proposes to the user but that are not interacted with. The process is depicted in Figure \ref{fig:personalization}. The adaptation of the user model $MLP_u$ requires a few steps of SGD minimizing the triplet loss
\begin{align}
    \mathcal{L}_{tri} = \sum_{(h_{p+}^{CNN},\ h_{p-}^{CNN}) \in B_u} &
    \max \big( 0, \|MLP_u(h_{wc}) - MLP_u(h_{p+}^{CNN})\|_2^2 \notag \\
    &\quad - \|MLP_u(h_{wc}) - MLP_u(h_{p-}^{CNN})\|_2^2 + \epsilon \big),
    \label{eq:triplet_loss}
\end{align}
where the hyper-parameter $\epsilon$, also named as ``margin'', has the effect of enforcing a minimum separation between positive and negative pairs in the embedding space preventing trivial solutions. 
The number of SGD update steps reflects the keenness to be pushing products closer. For instance, if all interactions in a batch are just click events, we may want to avoid the model changing too much; on the opposite, if all interactions are purchases we would like the model to project these articles very close together. 

The underlying intuition of our method is that we are dynamically re-organizing the representation of products, particularly those unknown by the user, making it more likely to suggest those that will be appreciated, and doing so through a very efficient and scalable approach.

\subsection{Technical details}

In Table \ref{tab:implementation} we report the computational and memory requirements to operate the system. The HGNN pre-training requires only a few hours of standard GPU computation; the response time for the recommendation part (1.5 ms) is compliant with our real-time objective, as well as the maximum memory occupancy of a few megabytes per user enables optimal scaling capability. It is fundamental to note that these phases are parallel and independent of each other; hence, the latency for the recommendations is unaffected by the time required for training and adapting the MLP. Additional details of how to maintain a constant memory occupation are discussed in Appendix~\ref{sec:sampling}. For an example of an instantiation of the system, refer to Appendix~\ref{sec:instantiation}. We performed an exhaustive grid search to optimize hyperparameters for both the HGNN and the online personalization component, as described in the Appendix~\ref{sec:model_selection}.

\begin{table}[t]
\footnotesize
\renewcommand{\arraystretch}{1.2}
\caption{Range of frequency, time required, and memory required for each of the main parts of the system. The actual values depend on hyper-parameters such as the number of SGD steps and the batch size.\protect\footnotemark}
\centering
\begin{tabular}{lccc}
    \toprule
    & \makecell{\textbf{HGNN-MLP} \\ \textbf{training (global)}} 
    & \makecell{\textbf{MLP adaptation} \\ \textbf{(per-user)}} 
    & \makecell{\textbf{Recommendation} \\ \textbf{(per-user)}} \\
    \midrule
    Frequency & 1–4 times/month & Every 1–9 interactions & At each interaction \\
    Time      & 1.5 h (GPU)     & 3–150 ms (CPU)        & 1.5 ms (CPU) \\
    Memory    & 16 GB (GPU)     & 2 MB                   & 700 KB \\
    \bottomrule
\end{tabular}
\label{tab:implementation}
\end{table}

\footnotetext{Hardware configuration: Nvidia Tesla T4 16 GB GPU, AMD Epyc 16-core CPU, 32 GB RAM.}

The framework is implemented in Python using Pytorch and Pytorch Geometric as main libraries.

\section{Results}
\label{sec:results}

We assess our approach on a proprietary dataset, referred to as the “e-commerce dataset.” For the sake of reproducibility, we also tested our model on a publicly available dataset released on Kaggle by H\&M \cite{kaggle_hm_dataset} for a global competition about personalization. The public dataset spans two years: from September 2018 to September 2020. It has 31 million transactions made by 1.3 million different users on 105,000 different articles; for each of those an image is available.

We simulate a realistic scenario by taking a temporal-based split, dividing the (simulated) past from the (simulated) future. The ``simulated past'' data are again split into train and validation in a user-based fashion. The training graph and the validation graph have in general different nodes and different edges since the users in the validation set may have interacted with a different set of articles compared with the ones in the training set. We used one week of data (simulated past) for training the HGNN and the following days (simulated future) for assessing recommendation and personalization in different scenarios: (i) day-by-day cold-start where at the end of each day the users' personalized spaces are deleted, (ii) multiple-weeks personalization, where the personalized embedding spaces are continuously adapted over time until the end of the tests. 
Three random weeks of data are sampled and used for pre-training (HGNN training) and for each of them, the following days are used for test (recommendations with continuous adaptation). The metrics used to assess the quality of the model are Precision, Recall, and F1 score. After each user's interaction, we predict $K$ articles the user will buy, and the ground truth set is composed of the next $T$ articles the user will actually buy. The values for $K$ and $T$ are set to $10$ and $12$ respectively, as it is common practice in these scenarios. 

\begin{table}[t]
\scriptsize
\caption{Performance comparison on the e-commerce dataset. All metrics (Precision, Recall, F1-score) are multiplied by $10^4$ for readability. Standard deviations are reported. Best results are in {\textbf{bold}}. The last two columns show the F1-score improvement of each model relative to LightGCN and PinSage respectively.}
\centering
\renewcommand{\arraystretch}{1.2}
\begin{tabular}{lccccc} 
    \toprule
    \textbf{Model / Baseline} & \textbf{Precision} & \textbf{Recall} & \textbf{F1-score} & \makecell[c]{\textbf{Improv. vs} \\ \textbf{LightGCN}} & \makecell[c]{\textbf{Improv. vs} \\ \textbf{PinSage}} \\
    \midrule
    Random Baseline & \(61\) & \(53\) & \(58\) & \(-74.7\%\) & \(-80.2\%\) \\ 
    Last-$K$ Baseline & \(174\) & \(145\) & \(158\) & \(-31.0\%\) & \(-46.1\%\) \\ 
    LightGCN & \(\mathbf{540}{\scriptstyle \pm 39}\) & \(147{\scriptstyle \pm 6}\) & \(229{\scriptstyle \pm 4}\) & - & \(-21.8\%\) \\
    PinSage  & \(352{\scriptstyle \pm 21}\) & \(246{\scriptstyle \pm 17}\) & \(293{\scriptstyle \pm 24}\) & \(+28.0\%\) & - \\
    \midrule
    HGNN + Daily personalization & \(371{\scriptstyle \pm 11}\) & \(306{\scriptstyle \pm 8}\) & \(336{\scriptstyle \pm 9}\) & \(+46.7\%\) & \(+14.7\%\) \\
    HGNN + 1 week personalization & \({448}{\scriptstyle \pm 61}\) & \(\mathbf{373}{\scriptstyle \pm 51}\) & \(\mathbf{407}{\scriptstyle \pm 56}\) & \(+77.7\%\) & \(+38.9\%\) \\
    HGNN + 2 weeks personalization & \(406{\scriptstyle \pm 45}\) & \(339{\scriptstyle \pm 37}\) & \(369{\scriptstyle \pm 41}\) & \(+61.1\%\) & \(+25.9\%\) \\
    HGNN + 3 weeks personalization & \(389{\scriptstyle \pm 45}\) & \(324{\scriptstyle \pm 38}\) & \(354{\scriptstyle \pm 41}\) & \(+54.6\%\) & \(+20.8\%\) \\
    \bottomrule
\end{tabular}
\label{tab:results_d2}
\end{table}

The results on the e-commerce dataset are shown in Table \ref{tab:results_d2}. The last-$K$ baseline uses as a prediction for the next $K$ purchases the last $K$ interacted products. We also compared our approach with two of the most used recommendations systems in the literature: PinSage~\cite{ying2018graph} and LightGCN~\cite{he2020lightgcn}. Adapting these models to our use case is not straightforward, we discuss in details about the baselines in Appendix~\ref{sec:baselines}. Table~\ref{tab:results_d2} shows that our model, after one week of adaptation, achieves the best results in F1 score. The high Precision and low Recall of LightGCN means that the model is accurate and conservative in its positive predictions but misses many true positive; its ability to find all relevant items (Recall) might be limited if those relevant items less connected in the interaction graph, or primarily discoverable through content features it doesn't use. In general a recommendation system with high Precision and low Recall is not optimal, as we discuss in more detail in Appendix~\ref{sec:baselines}.
It is worth noting that in our current experimental setting, adaptation is evaluated globally, with all user interactions constrained between fixed start and end dates. We hypothesize that adopting a more fine-grained approach, such as individually adapting for one week starting from each user's first interaction, could further improve the resulting metrics.

\begin{table}[t]
\caption{Performance comparison on the public dataset. All metrics (Precision, Recall, F1-score) are multiplied by $10^4$ for readability. Standard deviations are reported. Best results are in {\textbf{bold}}. The 4th column is the F1 improvement with respect to the pre-trained model (no personalization) and the 5th column is the F1 improvement with respect to the best public solution.}
\centering
\scriptsize
\renewcommand{\arraystretch}{1.2}
\begin{tabular}{lccccc}
    \toprule
    \textbf{Model / Baseline} & \textbf{Precision} & \textbf{Recall} & \textbf{F1} & \makecell[c]{\textbf{Improv.} \\ \textbf{no pers.}} & \makecell[c]{\textbf{Improv.} \\ \textbf{best}} \\
    \midrule
    Random baseline & \(72\) & \(59\) & \(62\) & \(-77.5\%\) & \(-76.3\%\) \\
    Best public solution \cite{jacobcp_handm_helpers} & \(289\) & \(241\) & \(262\) & \(-5.1\%\) & - \\
    \midrule
    HGNN No personalization & \(303{\scriptstyle \pm 26}\) & \(253{\scriptstyle \pm 22}\) & \(276{\scriptstyle \pm 23}\) & - & \(+5.3\%\) \\
    HGNN + Daily personalization & \(309{\scriptstyle \pm 24}\) & \(258{\scriptstyle \pm 19}\) & \(282{\scriptstyle \pm 20}\) & \(+2.2\%\) & \(+7.6\%\) \\
    HGNN + 1 week personalization & \(\mathbf{313}{\scriptstyle \pm 20}\) & \(260{\scriptstyle \pm 18}\) & \(284{\scriptstyle \pm 18}\) & \(+2.9\%\) & \(+8.4\%\) \\
    {HGNN + 2 weeks personalization} & \({312}{\scriptstyle \pm 22}\) & \(\mathbf{285}{\scriptstyle \pm 15}\) & \(\mathbf{298}{\scriptstyle \pm 20}\) & {\(+8.0\%\)} & {\(+13.7\%\)} \\
    HGNN + 3 weeks personalization & \(313{\scriptstyle \pm 22}\) & \(260{\scriptstyle \pm 18}\) & \(284{\scriptstyle \pm 20}\) & \(+2.9\%\) & \(+8.4\%\) \\
    \bottomrule
\end{tabular}
\label{tab:results_d1}
\end{table}

In Table~\ref{tab:results_d1}, we present the results obtained on the public dataset. All of our adapted models surpass the best publicly available solution~\cite{jacobcp_handm_helpers}. The improvements over this baseline is more modest compared to the e-commerce dataset due to the limitations of this public dataset, which contains only purchase events, lacking richer user interaction signals such as add-to-cart, favorite, and click events. Additionally, it is important to highlight that the original Kaggle competition involved a different and simpler task: predicting a single set of $K$ future purchases for each user based on their entire purchase history, without any time or memory constraints to compute the prediction, giving a strong advantage to non-real-time solutions.

\subsection{Ablation Study}
\label{sec:ablation_study}

\begin{table}[t]
\caption{Ablation study results for different configurations of our recommendation system on the e-commerce dataset. All metrics (Precision, Recall, F1-score) are multiplied by $10^4$ for readability. Standard deviations are reported. Best results are in {\textbf{bold}}. The ``Improv./Degrad.'' column shows the percentage change in F1-score relative to our ``Complete'' model HGNN + 1 week Personalization. The last two columns show the F1-score improvement relative to LightGCN and PinSage.}
\centering
\scriptsize 
\renewcommand{\arraystretch}{1.3}
\begin{tabular}{lcccccc} 
    \toprule
    \makecell[l]{\textbf{Model} \\ \textbf{Configuration}} & \textbf{Precision} & \textbf{Recall} & \textbf{F1-score} & \makecell[c]{\textbf{Improv./} \\ \textbf{Degrad. vs} \\ \textbf{Complete}} & \makecell[c]{\textbf{Improv. vs} \\ \textbf{LightGCN}} & \makecell[c]{\textbf{Improv. vs} \\ \textbf{PinSage}} \\
    \midrule
    LightGCN & \(\mathbf{540}{\scriptstyle \pm 39}\) & \(147{\scriptstyle \pm 6}\) & \(229{\scriptstyle \pm 4}\) & \(-43.7\%\) & - & \(-21.8\%\) \\
    PinSage  & \(352{\scriptstyle \pm 21}\) & \(246{\scriptstyle \pm 17}\) & \(293{\scriptstyle \pm 24}\) & \(-28.0\%\) & \(+28.0\%\) & - \\
    \midrule
    \makecell[l]{HGNN + 1 week \\ Personalization} & \(448{\scriptstyle \pm 61}\) & \(\mathbf{373}{\scriptstyle \pm 51}\) & \(\mathbf{407}{\scriptstyle \pm 56}\) & - & \(+77.7\%\) & \(+38.9\%\) \\
    \midrule
    No Personalization & \(336{\scriptstyle \pm 15}\) & \(280{\scriptstyle \pm 12}\) & \(305{\scriptstyle \pm 14}\) & \(-25.1\%\) & \(+33.2\%\) & \(+4.1\%\) \\
    No Pre-training & \(396{\scriptstyle \pm 17}\) & \(330{\scriptstyle \pm 21}\) & \(360{\scriptstyle \pm 13}\) & \(-11.6\%\) & \(+57.2\%\) & \(+22.9\%\) \\
    \makecell[l]{No Pre-training and \\ No Personalization} & \(224{\scriptstyle \pm 7}\) & \(196{\scriptstyle \pm 22}\) & \(209{\scriptstyle \pm 19}\) & \(-48.7\%\) & \(-8.7\%\) & \(-28.7\%\) \\
    \bottomrule
\end{tabular}
\label{tab:ablation_results}
\end{table}

To assess the contribution of each primary component within our proposed recommendation framework, we conducted a series of ablation studies. 
These experiments systematically remove parts of our architecture to isolate their impact on recommendation performance. 
We evaluate these ablated models on the e-commerce dataset using Precision, Recall, and F1-score as described in Section~\ref{sec:results}. In Table~\ref{tab:ablation_results} the following configurations are evaluated:

\textbf{No personalization (Cold-Start Performance).} This configuration removes the continual adaptation mechanism (i.e., the Personal MLP is not fine-tuned during the user's session). Recommendations are generated using the user's session history (via EMA) projected by the global, pre-trained Student MLP.

\textbf{No Pre-training (Continual personalization Only).} In this setup, the personalized MLPs are initialized without any knowledge distillation from a pre-trained HGNN. The Personal MLPs are then continually adapted based on user interactions as in the full model.

\textbf{No Pre-training \& No personalization (CNN-EMA Baseline).} This represents a minimal pipeline, serving as a strong baseline. It directly uses the raw CNN image embeddings of items. The user's current interest is represented by EMA of the CNN embeddings of items they have interacted with in their session. This configuration ablates both the graph-based pre-training and the adaptive MLP personalization.

The results in Table~\ref{tab:ablation_results} highlight the need of both the HGNN pre-train and the continuous adaptation to user's preferences to achieves the best F1 score. Note that the Cold-Start (No Personalization) performance of our model is comparable to PinSage; in fact, in both cases, we have a pre-train phase via GNN and an EMA-style recommendation setting. Our model outperforms PinSage even without personalization phase likely due to the fact that we explicitly take into account multiple interaction types. Finally, the fact that LightGCN (that does not consider visual information) is surpassed by the simple CNN-EMA Baseline in terms of Recall suggests that in this recommendations scenario visual appearances are in general more important than global interaction patterns.

\section{Conclusions}
We proposed a novel scalable recommendation system exploiting a graph-based representation of historical interactions and continual adaptation of lightweight distilled models personalized to each user. Our pre-personalization model reaches competitive performance also for cold-start users, by exploiting the richness of representation of historical interactions. Knowledge distillation effectively transfers information to smaller models making the system faster and more memory-efficient, enabling real-time personalization for each user. The recommendation module is distance-based, which means that the latency is in the order of a few milliseconds and it is independent from the update time of the personalized models, which, however, requires only a few steps of SGD on a shallow MLP. Results show that the continuous adaptation is effective and enables us to perform more accurate predictions on future purchases than the best available solutions.
Future work includes experimenting with a replay memory for each user to transfer the knowledge after the model is discarded. In fact, results show that the best life span for the personalized models is 1 or 2 weeks, after which the performance begins to degrade. An increased stability, for example, a higher regularization could increase the service life of the models at the price of a lower plasticity (and possibly a worse performance in the short term). A replay memory would also allow seasonal replay, for example when the summer season arrives we could use the navigation data from the same period of the past years to recommend products from a new collection that match the the user's personal style.

\bibliographystyle{splncs04}
\bibliography{bib}

\newpage
\appendix

\section{Additional technical details}

\subsection{Sampling}
\label{sec:sampling}

The graph can have tens of thousands of nodes and millions of edges, is therefore impractical to compute all the nodes' embedding in a single pass due to GPU memory constraints. In this regard, inspired by Hamilton et al. \cite{hamilton2017inductive}, we follow a sampling approach to build graph batches, in addition we also experimented a weighted sampling procedure based on the edges' weights described in Section \ref{sec:methods}:

\begin{enumerate}
	\item The set of nodes is partitioned in batches such that each batch has \textit{batch\_size} nodes.
	\item Starting from each node in the batch, iteratively sample some edges giving more probability to edges with higher weighs.  
    \item Add the nodes to the other end of the selected edges (and the edges themselves) to the graph batch. 
\end{enumerate}

The number of iterations and the number of edges to add at each iteration is described by the \textit{num\textunderscore neighbor} list, for example the list $[3, 2]$ indicates to sample three edges for each node in the batch, and for each induced node sample two additional edges, and so add the two new nodes to the batch.

Only the edges used for the sampling process are added to the final batch graph, if there are other edges joining nodes in the batch, these are not considered in order to avoid an uncontrolled exponential increase of the edges.

At the end of the sampling procedure, the maximum number of edges in the batch $E_b$ will be the product of the elements in the \textit{num\_neighbor} list and the maximum number of nodes will be $|E_b| * batch\_size$. Consequently, the adopted batch sampling strategy ensures that we have a constant predefined memory occupation whatever is the size of the graph.

\subsection{Instantiation}
\label{sec:instantiation}

One possible instantiation of the system is the following:

\begin{itemize}
	\item The HGNN-MLP training is done at the end of each week with data from the week just ended. It will require 1.5 h but in the meantime the system continues working with the old MLPs instantiated for each user.
	\item When the HGNN-MLP training is complete, the new MLP is copied to all the users and the old ones are discarded.
	\item The user continues to navigate the website, every 5 interactions the MLP adaptation starts and lasts about 100ms, in the meantime the user continues receiving recommendations with the not yet updated MLP model\footnote{It is really rare that a new interaction arrives in less than 100 ms.}.
	For each user, there is always one available MLP model. Since the role of the MLP is to project the CNN embeddings in a different space, this can be done in batch just after the MLP adaptation is complete (the space is modified), hence the MLP is used once every adaptation, and the embedding produced are stored in memory.
	\item The computation of the recommendation is a K-nearest-neighborhood from the user's representation. Our implementation is a simple sorting-based one and it respects the specifics required. An advanced data structure could further reduce the latency but would require a pre-processing phase.
\end{itemize}

\section{Model selection}
\label{sec:model_selection}

We performed an exhaustive grid search to optimize hyperparameters for both the HGNN and the online personalization component (Table \ref{tab:hyperparameters_combined}). The activation function was consistently set to ReLU across all experiments. Motivated by prior studies that indicate the superiority of Euclidean distance over cosine similarity in low-dimensional embedding spaces \cite{mukherjee2023reconciliation,tessari2024surpassing,ladd2020understanding}, we adopted Euclidean distance throughout.

For the HGNN, we instantiated Equation \ref{eq:gnn} using SAGEConv \cite{hamilton2017inductive}, with a fixed batch size of 128 and neighbor sampling configuration of \([8,8,8]\) to ensure constant memory usage, as discussed in Appendix \ref{sec:sampling}. Additionally, we employed early stopping with a patience of 5 epochs and set loss weighting to \(\gamma = [\gamma_1= 1, \gamma_2= 0.5, \gamma_3= 0.5, \gamma_4= 0.1]\).
 
All other hyperparameters listed in Tables \ref{tab:hyperparameters_combined} were optimized using grid search.

\begin{table}[h!]
\centering
\scriptsize
\caption{Hyper-parameters for the heterogeneous graph model (left) and online personalization (right).}
\label{tab:hyperparameters_combined}
\begin{minipage}{0.49\textwidth}
    \centering
    \begin{tabular}{ll}
        \toprule
        \multicolumn{2}{l}{\textbf{Heterogeneous Graph Model}} \\
        \midrule
        \textbf{Hyper-parameter} & \textbf{Values} \\
        \midrule
        structural layers & \(\begin{aligned}[t]
            &[\text{dim\_in}, 256, 128, 64],\\[-0.4em]
            &[\text{dim\_in}, 256, 256, 128, 64]
        \end{aligned}\) \\
        attribute layers & \(\begin{aligned}[t]
            &[\text{dim\_in}, 256, 128, 64],\\[-0.4em]
            &[\text{dim\_in}, 256, 256, 128, 64]
        \end{aligned}\) \\
        neighborhood agg. & \(sum, mean\) \\
        relation agg. & \(sum, mean\) \\
        weighted sampling & \(True, False\) \\
        learning rate & \(\begin{aligned}[t]
            &1e{-2}, 1e{-3}, 1e{-4},\\[-0.4em]
            &1e{-5}, 1e{-6}, 1e{-7}
        \end{aligned}\) \\
        weight decay & \(1e{-5}, 1e{-6}, 1e{-7}, 0\) \\
        margin & \(1, 100\) \\
        \bottomrule
    \end{tabular}
\end{minipage}%
\hfill
\begin{minipage}{0.49\textwidth}
    \centering
    \begin{tabular}{ll}
        \toprule
        \multicolumn{2}{l}{\textbf{Online Personalization}} \\
        \midrule
        \textbf{Hyper-parameter} & \textbf{Values} \\
        \midrule
        SGD steps & \(\begin{aligned}[t]
            &1, 2, 3, 4, 5, 10,\\[-0.4em]
            &20, 35, 50, 65, 80
        \end{aligned}\) \\
        learning rate & \(\begin{aligned}[t]
            &1e{-2}, 1e{-3}, 1e{-4},\\[-0.4em]
            &1e{-5}, 1e{-6}
        \end{aligned}\) \\
        weight decay & \(1e{-5}, 1e{-6}, 1e{-7}, 0\) \\
        batch size & \(1, 2, 3, 5, 7, 9\) \\
        margin & \(1, 100, \infty\) \\
        alpha & \(\begin{aligned}[t]
            &0.1, 0.2, 0.3, 0.4, 0.5,\\[-0.4em]
            &0.6, 0.7, 0.8, 0.9, 1
        \end{aligned}\) \\
        \bottomrule
    \end{tabular}
\end{minipage}
\end{table}

\section{Baselines}
\label{sec:baselines}

The last-K baseline is important because a poor recommendation system could simply suggest to buy all the previously clicked or added to cart articles. This achieves a moderately good performance in terms of Accuracy when simulating the streaming with website logs, in fact, a purchased product has certainly been clicked and added to the cart in the past, and the recommendation system could learn to predict as future purchase only the items added to the cart. However, in real-world usage, this behavior will not increase the users' satisfaction and the number of purchases in the platform.

Directly comparing our proposed method with real-time personalization against models like LightGCN \cite{he2020lightgcn} and PinSage \cite{ying2018graph} in their canonical forms presents inherent challenges due to fundamental differences in their architectural designs and primary objectives. 
LightGCN, as detailed by He et al. (2020), is explicitly designed for pure collaborative filtering, and thus does not inherently leverage item content features like image embeddings. This makes a direct comparison difficult when item content is a significant aspect of the recommendation task, as it is in our fashion e-commerce context.

PinSage, on the other hand (Ying et al., 2018), while capable of incorporating content features to learn powerful static item embeddings from large-scale item-item or item-board graphs, typically employs these embeddings in a downstream recommendation phase. Its core architecture focuses on generating these item representations, and user personalization is often achieved by querying these static embeddings based on recent user activity, rather than through real-time adaptation of a user-specific model component as in our approach. The original PinSage paper, for instance, details a method for learning these item embeddings primarily for item-to-item recommendation scenarios.

To address these disparities and facilitate a more equitable comparison for the purpose of evaluating the quality of learned item representations in a dynamic, session-based context, we standardizes the recommendation generation and evaluation protocol. 
Specifically, after training the models to learn item embeddings from an item-item graph constructed from user interaction sequences (and incorporating item image features as node attributes in PinSage), we evaluate these learned embeddings in a manner analogous to our main proposal's inference stage:
\begin{enumerate}
    \item The user's evolving interest within a session is represented by an Exponential Moving Average (EMA) of the learned item embeddings corresponding to the items interacted with so far in that session.
    \item Recommendations are generated by performing a $K$-Nearest Neighbor (K-NN) search against this session-specific EMA user profile vector within the global space of learned item embeddings.
    \item The same session-based Precision, Recall, and F1-score metrics (comparing against the next $T$ items, as detailed in Section~\ref{sec:results}) are then applied.
\end{enumerate}

\end{document}